\documentclass[%
 reprint,
 amsmath,amssymb,
 aps,
]{revtex4-1}

\usepackage{graphicx}
\usepackage{dcolumn}
\usepackage{bm}

\usepackage{amssymb} 
\usepackage{braket}
\usepackage{mathcomp}
\usepackage{amsmath}

\begin{document}


\title{Quantum gates implementation by X-ray single-photons around rotating black holes}

\author{Ovidiu Racorean}
 \email{ovidiu.racorean@mfinante.ro}
\affiliation{%
General Direction of Information Technology, Bucharest, Romania.\\
}%




\date{\today}

\begin{abstract}
The curvature and twisting of spacetime rotate the angle of polarization and imprint orbital angular momentum to photons emitted by the accretion disk near rotating black holes. Considering polarization and orbital angular momentum as two degrees of freedom of single-photons that can encode quantum information, we emphasize that the particular shape of spacetime around rotating black holes implements quantum gates and simple quantum circuits. Consequently, we demonstrate the implementation of some elementary quantum gates, like Hadamard or C-NOT, and simple quantum circuits, like Bell states, with photons in the presence of spinning black holes. Detection and measurement of quantum information encoded in photons emitted in the accretion disk around rotating black holes may be performed by actual quantum information technology.
\begin{description}
\item[PACS numbers]
 04.70.Dy, 03.67.Bg, 42.50.Ex, 95.30.Gv
\end{description}
\end{abstract}

\pacs{Valid PACS appear here}
\maketitle


\section{\label{sec:level1}Introduction}

Classical work related to determine the polarization of X-ray photons emitted near RBH considers the Newtonian limit, which consists, for symmetry reasons, on horizontal polarization, parallel to the disk and vertical polarization, perpendicular to the disk \cite{chan}. However, the polarization of photons is determined by local processes in the source and by general relativistic effects within the source and during the propagation to the observer. In recent numerical simulations  \cite{sch},  \cite{con},  \cite{conn},  \cite{cun},  \cite{agol} the role of general relativistic effects is strongly emphasized and determines the angle of polarization to acquire values in the interval $[-〖90〗^o,〖90〗^o]$. It was pointed out that the radiation deflected by the strong gravitational field near RBH and scattered by the accretion disk toward a distant observer, as returning radiation,suffers larger rotations of the polarization vector. 

We chose here to characterize the polarization of the radiation emitted near the RBH by the density matrix, a less explored \cite{mcm1}, \cite{mcm2}, \cite{fan}, \cite{acq}, but equivalent approach\cite{fal}, having in mind to take further advantages of the lately development of quantum information theory. The density matrix completely determines the polarization states of photons \cite{gam}, \cite{gil} and directly points out to polarization as a degree of freedom of photons that can encode quantum information, in the form of one qubit. Accordingly, we consider horizontal and vertical polarization of photons as encoding the qubit state \cite{shw} , \cite{gun} and we infer polarization as a two-level quantum system that spans the two-dimensional Hilbert space.

Polarization is not the only degree of freedom the single-photons acquire near RBH. It was advocated \cite{tam}, \cite{yang} that the twisting of spacetime, reflected in the dragging frame effect near RBH, imprints orbital angular momentum (OAM) to  radiation beams emitted by the accretion disk. Numerical simulations have shown a spectrum of OAM that widens as the BH spins faster. The OAM flips to wider modes for photons emitted in the innermost regions of the disk that are deflected and scattered to a distant observer as returning radiation. Traditionally, in quantum information theory \cite{sas}, \cite{pel} the parity of OAM modes of photons may encode quantum information, as a single qubit.

In this context, to fully characterize the states of photons emitted by accretion disk, a distant observer must measure both degrees of freedom acquired by photons traveling in the spacetime deformed by the presence of RBH – polarization and OAM. Considering the quantum information encoded in the two degrees of freedom of photon, as two qubits, \cite{den}, \cite{fio}, \cite{per}, we determine here how the particular shape of spacetime near RBH manipulates the quantum information and implements elementary quantum gates and simple quantum circuits, like Bell states.

We demonstrate the implementation of the polarization Hadamard gate \cite{hei}, \cite{cres} NOT gate and CNOT gate by general relativistic effects probed by photons near RBH. Implementation of CNOT quantum gate is important since it provides evidence that the shape of spacetime entangles the two degrees of freedom of photons. Accordingly, photons in diagonal and anti-diagonal states could realize the all four Bell states.   

It is expected that the photons detected from RBH sources should carry a spectrum of quantum information, from photons in separable states to nonmaximally entangled states photons \cite{whi}, \cite{bar} and photons carrying Bell states \cite{kha}.  

The future measurements of the Bell states encoded in single-photons emitted near RBH are promising since the technology required by such a process is already in use.

\section{\label{sec:level2}	Polarization of X-ray photons near spinning black holes}

The main source of information on the efforts to unveil some of the black holes hidden characteristics, such as the speed of spinning, is the polarization of radiation emitted by the accretion disk. The detection of  the polarization vector rotations of photons may be a strong evidence of general relativistic effects that manifest near RBH. Rotations of polarization angle of radiation emitted by accretion disks in the thermal state, have been shown in numerical simulations  \cite{sch},  \cite{con},  \cite{conn},  \cite{cun},  \cite{agol}, assuming the simplest model of a geometrically thin, optically thick, steady-state accretion disk, aligned with the BH spin axis.

Due to its simplicity, the Stokes parameters formalism is preferred in numerical simulation models in order to characterize the polarization angle of radiation emitted by the accreting black holes. Considering that the polarization is induced by Compton scattering, such that no circular polarization is present, the polarization angle is derived from the Stoke parameters in the following manner:

\begin{equation}
tan2\chi=\frac {U}{Q}  
\end{equation}

where $U$, $Q$ are the Stokes parameters related to linear polarization of photons, $\chi$ is the polarization angle. 

The polarization angle of the radiation emitted near RBH is sensitive to the black hole spin parameter ($a$) and the angle of inclination ($i$) of the accretion disk in relation to the distant observer. The angle of inclination is not relevant here and we assigned it a constant value  $i=〖45〗^0$ , inferring further only the spin parameter of the black hole. 

Numerical simulations \cite{sch}, \cite{cun} pointed out, in the context of the standard disk model, that the angle of polarization shows large variations with energy, corresponding to emission of different energy photons from different regions of the disk. Thus, it was shown that the radiation emitted near the outer region of the accretion disk, at low energies, is horizontally polarized, parallel to the plane of the disk, with the angle of polarization, $\chi=0^o$.

Accordingly, the radiation originating from the innermost region of the disk, probing the strong gravitational field near RBH is deflected back to the disk and scattered to the distant observer as returning radiation. The scattered radiation is vertically polarized, with the polarization angle perpendicular, $\chi=±〖90〗^o$, to the plane of the disk.

In the context of rapidly spinning black holes ($a=0.98$), photons emitted between the outer and innermost regions of the accretion disk, in the transition region, as direct radiation, suffer large rotations of the polarization vector toward positive values $\chi>0$, peaking at energies around 1 keV with diagonal polarization, $\chi=〖45〗^o$. Consequently, for the returning radiation, at high energies, it is prescribed a rotation of the polarization vector in the clockwise direction, giving a negative angle of polarization, $\chi<0$, that may reaches the anti-diagonal polarization, $\chi=-〖45〗^o$ . Slowly rotating black holes ($a=0.5$) and non-rotating BH ($a=0$) have shown reduced rotations of polarization angle and at energies above 10 keV.

The polarization of radiation can be characterized either by the Stokes parameters or by, a less considered approach \cite{mcm2}, \cite{fan}, \cite{acq} the 2x2 density matrix. The two approaches are equivalent as it was demonstrated in \cite{fal}. We consider here the density matrix representation of the polarization of radiation emitted near RBH, having in mind to speculate the close analogy with the quantum mechanics density matrix \cite{acq} and take advantages of the latest development of quantum information science.

We construct the density matrix via the coherency matrix starting from Stokes parameters of the photons emitted near rotating black holes. The coherency polarization matrix unitary contains all physical information about the polarization state of radiation. The polarization matrix, considering the absence of circular polarization is given by:

\begin{equation}
    \rho_{pol} = \frac{1}{2}\begin{bmatrix}I+Q & U  \\U & I-Q \end{bmatrix}
\end{equation}

Straightforward calculations considering \cite{gam}, \cite{gil}, lead to:

\begin{equation}
    \rho_{pol} = \begin{bmatrix}cos^2\chi & cos\chi  sin\chi \\cos\chi  sin\chi& sin^2\chi \end{bmatrix}
\end{equation}

The polarization matrix equals the density matrix of the photons source \cite{gam}, \cite{gil}, since density matrix is the unit trace scaling of the polarization matrix and $Tr(\rho_{pol} )=1$, such as for the rest of the present paper we will refer to $\rho_{pol}$ as the density matrix of polarization of photons.

The density matrix in Eq.(3) is Hermitian, and expresses the two-level orthogonal system of X-ray photons polarization, considering the polarization as the degree of freedom.

The states of the X-ray photons span the Hilbert ($H_{pol}$) two-dimensional space of polarization $\ket{\Phi_{pol}} \in H_{pol}$.
The density matrix in Eq.(3) corresponds to the general quantum state of X-ray photons emitted by the accretion disk near the rotating black hole: 

\begin{equation}
\ket{\Phi_{pol}}=cos\chi \ket{H}+sin\chi \ket{V}
\end{equation}

where $H$ and $V$ are horizontal and vertical polarization, respectively. We recognize in the Eq.(4) the general normalized representation of the qubit state \cite{den}, \cite{fio}. The polarization degree of freedom encodes quantum information, as a single qubit. Quantum information processing based on optical polarization is a flourishing field of research encompassing optical quantum computation and quantum communication. 

The state of photons is completely specified by knowing the polarization angle as it results from Eq.(4). Thus, the state of photons emitted at the outer edge of the disk, with the polarization angle $\chi=0^o$, is  $\ket{\Phi_{pol}}=\ket{H}$ , while the photons coming from the inner edge, with $\chi=〖90〗^o$, are in the state $\ket{\Phi_{pol}}=\ket{V}$.

An important observation we have to infer here is that at higher energies, in the inner regions of the disk, as we stated earlier, the angle of polarization rotates clockwise to negative values, $\chi<0$, such that the states:  

\begin{equation}
\ket{\Phi_{pol}}=cos\chi \ket{H}-sin\chi \ket{V}
\end{equation}

should also be considered. 

\section{\label{sec:level3}	Spacetime manipulation of polarization qubit}

The polarization of photons emitted by the BH’s accretion disk is determined by local processes in the source (Compton scattering) and by general relativistic effects within the source and during the propagation toward the observer. In the present approach we are interested in changes of the polarization generated by the strong general relativistic effects near RBH. Accordingly, we consider the polarization states of the photons, horizontal of vertical, due to the local processes in the source as input states, while the radiation beams escaping the BH’s influence towards the distant observer are considered as output states. 

Although, initially polarized horizontally or vertically, the strong general relativistic effects near RBH alter the state of photons by changing the polarization angle. The general relativistic effects allow polarization angles in any direction. More energetic photons coming from nearer the BH suffer larger general relativistic rotations with the angle of polarization taking values in the interval, $\chi \in [-〖90〗^o,〖90〗^o]$.

On the outer regions of the accretion disk, characterized by low energies, the polarization vector rotates to positive angles, $\chi>0$, for the direct radiation. Nevertheless, near the transition region where the effects of gravitational field start to manifest, the angle of polarization may reaches $〖45〗^o$, for a large portions of photons at energies around 1 keV, emitted close to rapidly spinning accreting black holes. 

The photons escaping the disk, initially horizontally polarized, as the input state, rotate the polarization angle during the propagation toward the distant observer, under the general relativistic effects into the output state characterized by diagonal polarization. Accordingly, measured by a distant observer, the output state of photons yields the diagonal polarization: 

\begin{equation}
\ket{\Phi_{pol}}=\frac{1}{\sqrt{2}}(\ket{H}+\ket{V})
\end{equation}

Close to the RBH, in the innermost regions of accretion disk, at high energies, polarization vector rotates in the clockwise direction, giving a negative angle of polarization, $\chi<0$. It is expected that initially vertically polarized photons, with high energies above the thermal peak, to shift the polarization state, as returning radiation, to the anti-diagonal polarization output state:

\begin{equation}
\ket{\Phi_{pol}}=\frac{1}{\sqrt{2}}(\ket{H}-\ket{V})
\end{equation}

The general relativistic effects, transform the initial horizontally polarized state of photons emitted near the transition region of the disk as:

\begin{equation}
\ket{H} \longrightarrow \frac{1}{\sqrt{2}}(\ket{H}+\ket{V})
\end{equation}

while the initial vertically polarized state as:

\begin{equation}
\ket{V} \longrightarrow \frac{1}{\sqrt{2}}(\ket{H}-\ket{V})
\end{equation}

The transformations of the states of photon’s polarization in Eq.(8) and Eq.(9) are well known and intensively studied lately in optics quantum information theory \cite{hei},\cite{cres}. These quantum transformations realize the elementary Hadamard gate \cite{hei}, with the qubit encoded in the polarization degree of freedom of photons. 

\section{\label{sec:level4}	OAM of photons near spinning black holes}

The polarization is not the only degree of freedom the photons emitted by the accretion disk near RBH may acquire. Performing numerical simulations of the radiation originated from the accretion disk of rotating black holes, Tamburini et al. \cite{tam}  observed nontrivial OAM generation and asymmetric spectra in terms of the LG-modes. Thus, the radiation beams from the accretion disk acquire independent azimuthal phase term $e^{i\ell\phi}$ possessing an orbital angular momentum of $\ell$ per photon, where $\ell$ is the integer topological charge. 

It was pointed out \cite{tam} that the OAM modes $\ell$ the frame dragging effect nearby spinning black holes imprint to X-ray radiation emitted by accretion disk are determined by the BH spin parameter ($a$) and the inclination angle of the disk ($i$) towards a remote observer. We maintain here for the inclination angle the same value $i=〖45〗^0$ as discussed in the case of X-ray polarization and analyze further only the BH spin parameter influence on the radiation coming from the accretion disk.

The spectrum of OAM modes acquired by direct radiation emitted near RBH that have two different values for the spin parameter, a relatively moderate spinning BH ($a=0.5$) to the left and a near-extreme rotating BH ($a=0.99$) to the right is depicted in Fig.1.

\begin{figure}
\includegraphics[width=8.6cm]{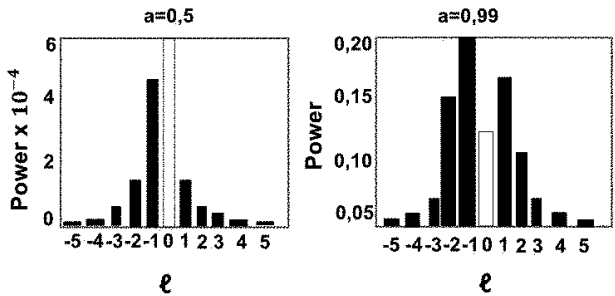}
\caption{\label{fig:OAM} OAM modes spectrum of X-ray photons emitted by the accretion disk of two different spin parameter black holes:  moderate spinning BH (left) and fast spinning BH (right).}
\end{figure}

We should infer from Fig.1 that the moderate spinning BHs ($a=0.5$) are characterized by a narrow spectrum of OAM modes dominated by zero OAM mode, with ($\ell=0$). The OAM spectrum reduces to zero LG mode ($\ell=0$) for Schwarzschild BH ($a=0$). 

On the other hand, rapidly spinning black holes ($a=0.99$) have the innermost region of the accretion disk closer to the black hole event horizon. X-ray photons emitted by the accretion disk are influenced stronger by the gravitational field because of the BH proximal vicinity and tends to flip toward wider OAM modes. As it can be noticed from Fig.1 the near-extreme spinning Black hole has a broader OAM spectrum with a significant reduction of zero LG mode.

It is worth being emphasized here that Tamburini et al. numerical simulations have shown that the major contribution to a widen OAM spectrum for the photons emitted by accretion disk, comes from the inner regions closer to the BH’s horizon, as returning radiation. Accordingly, photons flip the OAM modes parity more rapidly as the radiation originates closer to the RBH. 

The photons emitted from the outer regions the accretion disk possess narrow OAM modes ($\ell=0,1$), while photons emitted from the innermost regions, experiencing stronger gravitational field near the RBH flip to wider OAM modes ($\ell=1,2,3..$).

Although, numerical simulations conclude that the initially zero OAM mode ($\ell=0$ ) of photons flips to wider OAM modes is independent of the photon frequency, X-ray photons emitted at higher energies, closer to the BH, are likely to acquire wider OAM modes. OAM modes of X-ray photons are intensively sudied lately \cite{sas}, \cite{pel}.

Our interest in the OAM modes of photons emitted by accretion disks is motivated here by the capacity of photons OAM modes to form a degree of freedom that can encode quantum information of a single qubit. Traditionally, in quantum information theory \cite{den}, \cite{per}, the OAM spans an infinite dimensional Hilbert space, and can theoretically take values in the interval, $[-\infty,\infty]$. We restrict here the OAM to a two-dimensional Hilbert space by considering the parity of OAM modes. The even/odd ($\ell_e/\ell_o$) grouping of OAM parity spans a two-dimensional Hilbert space and may encode a single qubit \cite{per}. 

We can write the initial state of photons originating from the RBH’s accretion disk, considering the OAM parity as the degree of freedom to encode a qubit is $LG_0$ ($\ell=0$ ), hence the OAM even parity, $\ell_e$, as:

\begin{equation}
\ket{\Phi_{OAM}}=\ket{\ell_e}
\end{equation}

It is likely that photons emitted al larger radii of the disk, probing low gravitational field, to escape the RBH influence toward a distant observer in the state in Eq.(10). 

The photons emitted by the innermost regions of the disk, probing stronger twist of spacetime near RBH, flip the initial zero OAM to a wider mode, the even OAM parity, $\ell_e$ transforms to odd OAM parity, $\ell_o$.

\begin{equation}
\ket{\Phi_{OAM}}=\ket{\ell_o}
\end{equation}

Changes in the OAM modes to wider values, hence the flipping of the initial even/odd OAM parity to a final odd/even OAM parity, acts similarly to unitary transformation that implements a NOT gate \cite{per}. 

\section{\label{sec:level5}	Spacetime implementation of C-NOT quantum gate}

The local processes at the source (Compton scattering) polarize photons horizontally or vertically depending on the region of the disk they are emitted, and hence, depending on the photons energy. During the propagation to the distant observer, influenced by the general relativistic effects near the black holes, photons suffer rotation of polarization angle and flip of the OAM parity to wider values. To completely determine the state of X-ray photons emitted by the accretion disk, both of the two degrees of freedom, polarization and parity of OAM, must be considered.  Accordingly, when escaping the RBH influence toward the distant observer, the states of photons should be expressed as a bipartite quantum system, $H_{pol}\otimes H_{OAM}$, that spans a 4-dimensional Hilbert space: 

\begin{equation}
\ket{\Phi_{pol-OAM}} \in H_{pol}\otimes H_{OAM}
\end{equation}

The state of photons is represented by a tensor product that aggregates the two distinct degrees of freedom of photons, polarization and parity of OAM, in a composite system \cite{den} \cite{per}. We could consider the state of photons emitted by the accretion disk and reaching a distant observer in the general form:

\begin{equation}
\ket{\Phi_{pol-OAM}} = \ket{\Phi_{pol}}\otimes \ket{\Phi_{OAM}}
\end{equation}

The initial parity OAM mode of the photons polarized after the local processes at the source is $〖LG〗_0$, with $\ell=0$ , hence, maintaining the even/odd parity notation, $\ket{\Phi_{OAM}} =\ket\ell_e$  . Considering $\ket{\Phi_{pol}}$  and $\ket{\Phi_{OAM}}$ in the forms we stated earlier, the general state of photons in Eq.(13) yields:

\begin{equation}
\ket{\Phi_{pol-OAM}} = (cos\chi \ket{H}+sin\chi \ket{V})\otimes \ket{\ell_e}
\end{equation}

which, by dropping the tensor product notation, can be simplified to:

\begin{equation}
\ket{\Phi_{pol-OAM}} = cos\chi \ket{H}\ket{\ell_e}+sin\chi \ket{V}\ket{\ell_e}
\end{equation}

It is likely that photons emitted by the accretion disk closer to the black hole, probing more the strong gravitational field, to flip the OAM modes to wider values. For the X-ray emitted by the innermost regions there is a flip of the OAM mode to odd values, escaping toward a remote observer in the status:

\begin{equation}
\ket{\Psi_{pol-OAM}} = cos\chi \ket{H}\ket{\ell_o}+sin\chi \ket{V}\ket{\ell_o}
\end{equation}

We noted here the state of photons, $\ket{\Psi_{pol-OAM}}$ to maintain unitarily the notation in quantum information science.
We emphasize here that if the photons are emitted at larger radii it is expected to escape toward a distant observer, as direct radiation in the state in Eq.(15), while the photons coming from regions closer to the black hole, in the state in Eq.(16).

The radiation emitted by the accreting RBH near the innermost regions likely returns to the disk and is scattered toward the observer as returning radiation.  The returning radiation, strongly vertically polarized, probing more the general relativistic effects (twisting of spacetime) flips the OAM modes to wider values. The state of the photon in Eq.(15) taken as initial state of the returning radiation, will suffer changes in the OAM modes that affects the vertical polarization component. As result, the parity of the photons vertical component flips from even to odd OAM values and yields the output state in the form:

\begin{equation}
\ket{\Phi_{pol-OAM}} = cos\chi \ket{H}\ket{\ell_e}+sin\chi \ket{V}\ket{\ell_o}
\end{equation}

We apply similar reasoning to returning radiation initially in the state Eq.(16) and considering the shift of vertical component OAM parity from odd to even, we find the output state:   

\begin{equation}
\ket{\Psi_{pol-OAM}} = cos\chi \ket{H}\ket{\ell_o}+sin\chi \ket{V}\ket{\ell_e}
\end{equation}

These shifts of states are considered in the quantum information theory \cite{den}, \cite{fio}, \cite{per} as unitary transformations that implements C-NOT gates. The X-ray photons that returning radiation consist of are carrying the quantum information associated to nonmaximally entangled states of the two degrees of freedom, polarization and OAM parity, of X-ray photons.  

This result is important since it proves that the curvature and twist of spacetime near RBHs could literally entangle the two degrees of freedom of single-photons.

The incidence for such special output quantum states of photons emitted by accretion disks it is expected to be high, when measured by a distant observer for near extreme rotating black holes.

The nonmaximal entangled states in Eq.(17) and Eq.(18) are not the only output states the photons emitted by accretion disks can acquire traveling the distorted spacetime near RBH toward the distant observer. We should mention here the presence of two other nonmaximal entangled states that photons near RBH can encode, as returning radiation. Returning radiation, vertically polarized, have the polarization vector rotated to negative angle ($\chi<0$) under the general relativistic effects closer to RBH. As returning radiation, the OAM of the vertical component of polarization flips to wider values changing the party from the initial even/odd state to the odd/even output state. 

The photons emitted near the thermal peak, at larger radii, likely to possess initially even OAM party flip to odd OAM parity the vertical component of the polarization leading to the nonmaximally entangled state of the form:   

\begin{equation}
\ket{\Phi_{pol-OAM}} = cos\chi \ket{H}\ket{\ell_e}-sin\chi \ket{V}\ket{\ell_o}
\end{equation}

Finally, we prescribe for photons coming from the inner edge of the disk, with the initial odd OAM parity a flip to even OAM parity of the vertical component of polarization:  

\begin{equation}
\ket{\Psi_{pol-OAM}} = cos\chi \ket{H}\ket{\ell_o}+sin\chi \ket{V}\ket{\ell_e}
\end{equation}

All four nonmaximally entangled states are widely considered lately in the quantum information science \cite{whi}, \cite{bar}.

\section{\label{sec:level6}	Bell states creation in single-photons near rotating black holes}

It was advocated by Schnitman and Krolik \cite{sch} that the vast majority of X-ray photons above the peak energy, 1keV in the case of rapidly spinning black holes ($a=0.98$), in the transition zone and the innermost regions of the accretion disk, acquire either diagonal or anti-diagonal polarization. The photons in diagonal polarization states can escape the black hole influence toward the observer, as direct radiation, with the $LG_0$ initial OAM mode, $\ell=0$, hence even parity $\ell_e$ , in the output state: 

\begin{equation}
\ket{\Phi_{pol-OAM}} = \frac{1}{\sqrt{2}}(\ket{H}\ket{\ell_e}+ \ket{V}\ket{\ell_e})
\end{equation}

if photons are emitted at larger radii, or

\begin{equation}
\ket{\Psi_{pol-OAM}} = \frac{1}{\sqrt{2}}(\ket{H}\ket{\ell_o}+ \ket{V}\ket{\ell_o})
\end{equation}

in the case photons are emitted in the innermost regions of the disk, having the initial even OAM parity flipped to final odd OAM parity.

Photons emitted above the thermal peak are likely deflected by the strong gravitational field near black holes and scattered by the accretion disk to a remote observer as returning radiation. During the propagation from the disk through the regions outside the RBH’s influence, probing more the general relativistic effects, the vertically polarized component of returning radiation flips the initial even OAM mode ($\ell_e$) to the odd OAM mode,($\ell_o$). Returning radiation originating from the regions near the transition region leaves the RBH influence, in the output state: 

\begin{equation}
\ket{\Phi_{pol-OAM}} = \frac{1}{\sqrt{2}}(\ket{H}\ket{\ell_e}+ \ket{V}\ket{\ell_o})
\end{equation}

while the photons originating from the innermost regions of the disk in the output state:

\begin{equation}
\ket{\Psi_{pol-OAM}} = \frac{1}{\sqrt{2}}(\ket{H}\ket{\ell_o}+ \ket{V}\ket{\ell_e})
\end{equation}

These particular states of photons emitted by the accreting black holes are two Bell states, the most explored states in the quantum information science \cite{per}, \cite{kha}. Bell states express the maximal entangled states of the two degrees of freedom, polarization and OAM, that photon acquire in the vicinity of RBHs.

We infer here that the above Bell states are similar to a C-NOT applied to the input states determined by the diagonal states of photons. We conclude mentioning that the other two Bell states originate from the anti-diagonal polarization state of returning radiation. Accordingly, applying similar reasoning as in the case of C-NOT gate implementation, we infer that the photons in anti-diagonal polarization state are escaping the RBH influence in the state:

\begin{equation}
\ket{\Phi_{pol-OAM}} = \frac{1}{\sqrt{2}}(\ket{H}\ket{\ell_e}- \ket{V}\ket{\ell_o})
\end{equation}

if the returning radiation originated from near transition region, above the thermal peak, or in the state:

\begin{equation}
\ket{\Psi_{pol-OAM}} = \frac{1}{\sqrt{2}}(\ket{H}\ket{\ell_o}- \ket{V}\ket{\ell_e})
\end{equation}

if the photons are emitted closer to the inner edge of the accretion disk. We exemplify the creation of all four maximally entangled states of photons under the strong influence of the gravitational field near rapidly spinning black holes. 

\section{\label{sec:level7}	Detection of Bell states from photons emitted  near rotating black holes}

Originating from regions closer to the RBH, at higher energies, X-ray photons are the perfect candidates to detect and measure the quantum information encoded in their degrees of freedom acquired by probing the BH strong gravitational field. However, the processing of quantum information encoded in the photons degrees of freedom in the gravitational field of rotating black holes should be observed to all spectrum of light. Although, radiation with low energies likely may not encode Bell states, since it originates from very large radii, probing less the gravitational field of RBH, it certainly encodes nonmaximaly entangled states.  

We emphasize that the most probable quantum state to be detected and measured in low energy radiation, like visible light, is:  

\begin{equation}
\ket{\Phi_{pol-OAM}} = cos\chi \ket{H}\ket{0}+sin\chi \ket{V}\ket{1}
\end{equation}

since it originates from the larger radii of the accretion disk. 

The incidence for the other three states involving wider OAM modes, negative angle of polarization and diagonal states of the photons polarization may not be present in the quantum measurements of visible light coming from accreting black holes.

Detection and measurement of Bell states encoded in the degrees of freedom of radiation should be focused on very energetic photons, with frequencies in the band of X-ray or gamma-ray, originating from the inner edge of the accretion disk, closer to the RBH. Probing more the general relativistic effects in the vicinity of the black hole, X-ray photons may encode wider OAM values, may acquire larger rotations of polarization vector and negative angle of polarization.

Recalling the OAM spectrum in Fig.1 and considering a near-extreme spinning black holes we should notice here that the most probable values for the even OAM modes are $\ell_e= 0$ and $\ell_e= 2$, and the odd OAM modes $\ell_o=1$. Under these circumstances we emphasize that the Bell states which could be measured as being encoded in X-rays are:

\begin{equation}
\ket{\Phi_{pol-OAM}^\pm}=\frac{1}{\sqrt{2}}(\ket{H}\ket{0}\pm\ket{V}\ket{1})
\end{equation}

and,

\begin{equation}
\ket{\Psi_{pol-OAM}^\pm}=\frac{1}{\sqrt{2}}(\ket{H}\ket{1}\pm\ket{V}\ket{2})
\end{equation}

Also from Fig.1, we could infer that the incidence for $\ket{\Psi_{pol-OAM}^\pm}$ states should be substantially higher for rapidly spinning black holes while the probability to measure the $\ket{\Phi_{pol-OAM}^\pm}$ states should increase for slow rotating black holes. 
X-ray photons emitted near BH’s event horizon may experience more than one bending and scattering as the BH spins faster. Near-extreme spinning black holes may account for Bell states having wider OAM modes, like:

\begin{equation}
\ket{\Phi_{pol-OAM}} = \frac{1}{\sqrt{2}}(\ket{H}\ket{2}- \ket{V}\ket{3})
\end{equation}

Measurement of Bell states encoded by various degrees of freedom of photons is very common technique in quantum optics \cite{per}, \cite{kha}. Setups meant to measure the Bell states are common in every laboratory working in quantum information processing of qubits encoded in visible light. However, the processing of quantum information encoded in the degrees of freedom of photons with very short wavelengths, like X-ray or gamma-ray, is hard to implement since it necessitate sophisticate equipment required by their high energy.

Still processing quantum information with X-ray photons is a flourishing field of research lately. The extreme robustness of the quantum information encoded by the two degrees of freedom of X-ray photons, polarization and OAM, attracted a large interest \cite{shw}, \cite{gun} of researchers in this particular field of quantum information processing.  

We conclude by expressing our reasonable expectations that the near future space missions around the world would contain apparatus capable to detect Bell states in X-ray photons originating from spinning black holes across the universe.

\section{\label{sec:level8}	Conclusions}

We have shown in the present paper that strong general relativistic effects near rotating black holes may process quantum information encoded in the polarization and OAM degrees of freedom of X-ray photons emitted by the accretion disk.

The strong gravitational field distorts the spacetime nearby spinning black holes, forcing the spacetime around them to curve and twist. Extreme curvature and twist of spacetime affect the photons emitted by the accretion disk. Propagating throughout the particular shape of spacetime near RBH, the photons polarization angle suffers large rotations and acquire a spectrum of wide OAM modes. 

Considering a density matrix approach constructed from the Stokes parameter via the coherency matrix, we have shown that these two degrees of freedom  – polarization and OAM - the photons acquired near spinning black holes, encode two qubits, characterize the states of photons. 

We emphasized here that the particular shape of spacetime near spinning black holes implements unitary transformations over the two degrees of freedom of photons which act similarly to the elementary quantum gates. 
Transformations suffered by the photons due to general relativistic effects nearby RBH are similar to quantum information processing in quantum optics. We explained the setups of spacetime required to perform unitary transformations onto single-photons in order to implement quantum gates. 

We exemplified these quantum processes by some elementary quantum gates implementation. The elementary Hadamard gate, NOT gate, and CNOT gate were constructed with the help of photons, considering only the general relativistic effects nearby fast rotating black holes.  

We also have shown that more complex quantum information processes, such as the Bell states, can be constructed based on the spacetime shape near RBH. 

The incidence for such quantum states of X-ray photons should be high such as every single X-ray photon escaping the RBH influence to the observer should carry quantum information, to some extent.

\end{document}